\documentclass[aps,superscriptaddress,nofootinbib,preprintnumbers]{revtex4}

\usepackage{amsmath}
\usepackage{graphicx,tabularx}
 
\newcommand{\etal}{\textit{et al.}}

\def\ie{{\it i.e.\;}}
\def\eg{{\it e.g.\;}}

\newcommand{\gev}{~{\ensuremath\rm GeV}}

\newcommand{\fb}{~{\ensuremath\rm fb}}

\begin{document}
 
\date{\today}
 
\title{Maximum Significance at the LHC and Higgs Decays to Muons}

\preprint{MPP-2006-43}
 
\author{Kyle Cranmer}
\affiliation{Goldhaber Fellow,
             Brookhaven National Laboratory, USA}
\author{Tilman Plehn}
\affiliation{Heisenberg Fellow, Max Planck Institute for Physics,
             Munich, Germany \\
             and School of Physics, University of Edinburgh,
             Scotland}
 
\begin{abstract}
  We present a new way to define and compute the maximum significance
  achievable for signal and background processes at the LHC, using all
  available phase space information. As an example, we show that a
  light Higgs boson produced in weak--boson fusion with a subsequent
  decay into muons can be extracted from the backgrounds. The method,
  aimed at phenomenological studies, can be incorporated in
  parton--level event generators and accommodate parametric
  descriptions of detector effects for selected observables.
\end{abstract}
 
\maketitle
 

The Large Hadron Collider (LHC) will have a tremendous capacity to
search for new particles, such as the Standard Model Higgs boson or new
particles suggested by various scenarios for physics beyond the
Standard Model.  For such searches, it is important to asses the
experimental sensitivity, which requires a description of the
experimental search technique to isolate signal-rich data.
Traditionally, this has been accomplished by using \textit{ad hoc}
kinematic cuts.  At the parton--level this process of designing cuts
by hand to isolate signal-enhanced phase space regions (which emulates
the traditional experimental practice) is not necessary. In this paper
we present a new method of computing the statistical significance of a
hypothesized signal via direct integration of the likelihood ratio.
This technique does not require identification of powerful
discriminating variables or techniques to estimate probability density
functions from a discrete sample of events.  Instead, we compute the
likelihood ratio exactly over the full phase space, which implies that
this expected significance is an upper bound.  This maximal
significance indicates if a more detailed study with a full detector
simulation is warranted and provides a target significance to which
any experimental study can be compared.  \smallskip

To demonstrate the power of this method, we consider the production of
the Standard Model Higgs boson at the LHC via weak--boson fusion with
a subsequent decay to muons. Weak--boson fusion production of a Higgs
boson with a subsequent decay to tau leptons originally proposed in
Ref.~\cite{wbf_tau} has been firmly established by {\sc Atlas} and
{\sc CMS} as the main discovery channel for a light Higgs boson in the
Standard Model~\cite{karl} as well as in its supersymmetric
extension~\cite{wbf_mssm}.  While QCD effects can be a danger for most LHC
analyses, additional jet radiation turns into a useful tool in the
case of weak--boson fusion signals~\cite{wbf_minijet}.  Observation of the
same process with a decay to muons can experimentally confirm Yukawa
couplings and their scaling with the masses for non--third--generation
fermions.

The expected significance of a search for $H\to \mu\mu$ was estimated
for weak--boson fusion~\cite{wbf_muon} and gluon fusion~\cite{gg_muon}
production modes.  For a 120 GeV Higgs boson mass, the best kinematic
cuts found in Ref.~\cite{wbf_muon} result in a $1.8~\sigma$
significance. The authors of that analysis note that many observables
display additional discriminating power and suggest that neural
networks or other multivariate procedures could enhance the
sensitivity. Using our new method we find that the maximum possible
(target) significance for $H \to \mu\mu$ is much higher, \ie the cut
analysis can indeed be significantly improved.

\subsection{Neyman--Pearson Lemma}

Our approach is based on the Neyman--Pearson lemma: the likelihood
ratio is the most powerful variable or test statistic for a hypothesis
test between a simple (\ie having no free parameters) null hypothesis
--- background only --- and an alternate hypothesis --- signal plus
background~\cite{Kendall}. Maximum power is formally defined as the
minimum probability for a Type~II error (false negative) for a given
probability for a Type~I error (false positive). If we assume that the
signal--plus--background hypothesis is true, the most powerful method
has the lowest probability of mistaking the signal for a background
fluctuation. \smallskip

The Neyman--Pearson lemma is commonly used to claim optimality, but
these claims can be misleading.  The reason is that the probability
density function (pdf) of a multi-dimensional observable $x$ for a
given hypothesis is not experimentally known. Instead,
experimentalists typically use a discrete sample of events $\{x_i\}$
to approximately estimate the pdf~\cite{kernel}.  In practice, the
size of the sample limits the dimensionality of the pdf that can be
estimated to one or two dimensions, or it requires one to neglect
correlations among the observables -- both of which invalidate strict
claims of optimality.  In contrast, in phenomenology we can use the
parton--level transition amplitude for a process (at a given order in
perturbation theory) to exactly compute the pdf over the full phase
space.\smallskip

Two main ingredients are needed to calculate the distribution of the
likelihood ratio for the background--only and signal--plus--background
hypotheses. First, we have to evaluate identical sets of phase space
points for signal and background processes, which is not part of
standard Monte Carlo event generators. Secondly, we need to bootstrap
the likelihood ratio distribution for one event to the distribution
for a fixed luminosity including Poisson fluctuations. Both
ingredients are discussed in the next Section. We then consider an
example: a light Higgs boson produced via weak--boson fusion and
decaying to muons. To achieve a minimum level of realism, we generalize
our method to include experimental resolutions and detector effects.\bigskip

It should be noted that this work builds on several techniques used in
experimental analyses, but it extends that work and applies it in a
phenomenological context. For instance, the literature is replete with
measurement techniques that use - to varying degrees - the matrix
element to describe kinematic
distributions~\cite{earlyDLmethod,optimalObservables,LEPExample}.
A qualitative distinction of this work is that we are estimating the
sensitivity of a search for a hypothesized particle instead of
measuring a theoretical parameter with data (\eg the mass of the top
quark or the helicity of the $W$ boson). In particular, we are not 
trying to identify the maximum--likelihood estimator for a parameter
to be extracted from data. The process of evaluating the
likelihood of an event in real data is significantly different from
constructing hypothetical data sets, and this leads to significant
differences in the implementation of the algorithms (in particular,
the two main ingredients mentioned in the previous paragraph).  Our
approach to the incorporation of experimental resolutions is very
similar to the recent work at the Tevatron, generically referred to as
``matrix element method''~\cite{WHelicity,DLmethod, MEmethod}, and we
try to use similar notation and terminology to make the correspondence
clear. 
Furthermore, we build on the statistical techniques (\eg
Eqs.~\ref{eq:testStatistic}-\ref{eq:clb}) used in the LEP Higgs
working group~\cite{lepewwg}, which generally has not been matched
with the matrix element method.  \smallskip

In short, our method is a novel combination of the LEP statistical
formalism with parton--level transition amplitudes used to define and
compute a mathematically well defined maximum expected significance. Note that
we do not attempt to identify any powerful discriminating observables,
nor do we attempt to compute an observed significance based on
experimental data~\cite{bard}. Instead, we formulate and answer the
question: {\sl what is the maximum expected significance of a 
potential physics signal, \eg a Higgs decaying to muons?}

\subsection{Likelihood Ratio and Discovery Potential}

We first limit ourselves to a signal process and its irreducible
backgrounds, \ie signal and background processes with identical
degrees of freedom in the final state, distinguished by (kinematic)
distributions.  To compute the expected signal and background rates we
integrate the matrix elements squared over the phase space, with or
without (acceptance) cuts, using a Monte Carlo integration.  This
method probes the phase space with random numbers. Ideally, the
dimension of the random number vector $\vec{r}$ is given by the number
of degrees of freedom in the final--state momenta after all kinematic
constraints. The random number vector forms a (minimal) basis for all
final-state configurations. We can schematically write
\begin{equation}
\sigma_{\rm tot} 
= \int_{\rm cuts} d PS \; M_{PS} \; d\sigma_{PS} 
= \int_{\rm cuts} d \vec{r} \; M({\vec{r}}) \; d\sigma({\vec{r}})
\label{eq:int_first}
\end{equation}
where the phase space boundaries are included in the integral, and the
differential cross section $d\sigma(\vec{r})$ includes all phase space
factors and the Jacobian for transforming the integration to the
random--number basis. The integration over the parton distributions is
included in the phase space integral. The measurement function $M$ can
be used to include additional cuts or to incorporate event weights
(\eg particle identification efficiencies) as a function of any
observable. Removing unwanted parts of the phase space through cuts
on observable quantities consistently removes the contribution of these phase
space regions from all proccesses. Because the random
numbers parameterize the entire phase space, all potentially available
information about the process is included in the array of event
weights $(M \, d\sigma)({\vec{r}})$.
Note that this phase space
integration above is written assuming a simple cross section
expression $d\sigma$; however, it can be replaced with
any combination of differential cross sections which modern parton--level 
event generators predict.
\medskip

A cut analysis defines a signal--rich region bounded by upper and
lower limits on observables and then counts events in that region.
Ultimately, the variable that discriminates between signal and
background --- the test statistic --- is simply the number of events
observed in this region.  Predicting the expected number of background
events $b$ and signal events $s$ enables us to adjust the cut values which optimize the
experimental sensitivity. More
sophisticated techniques use multivariate algorithms, such as neural
networks, to define more complicated signal--like regions, but the
test statistic often remains unchanged. In all of these counting
analyses, the likelihood of observing $n$ events assuming the
background-only hypothesis is simply given by the Poisson distribution
${\rm Pois}(n|b)=e^{-b} \, b^n/n!$.\smallskip

There are extensions to this number counting, assuming we know the
distribution of a discriminating observable $x$ (which may be
multi-dimensional). We assume that for the background--only hypothesis
$H_0$ this distribution is $f_b(x)$, while for the
signal--plus--background hypothesis $H_1$ it is $f_{s+b}(x) =
\left[sf_s(x) + bf_b(x)\right]/(s+b)$ assuming no interference.
Following the Neyman-Pearson lemma, the most powerful test statistic
is the likelihood ratio for the entire experiment's data. The total
likelihood for the full--experiment observable ${\bf x}=\{x_j\}$ can
be factorized into the Poisson likelihood to observe $n$ events, and
the product of the individual event's likelihood $f(x_j)$:
\begin{alignat}{5}
Q({\bf x}) &= \frac{L({\bf x}|H_1)}{L({\bf x}|H_0)}
      = \frac{{\rm Pois}(n|s+b)  \; \prod_{j=1}^{n} f_{s+b}(x_j)} 
             {{\rm Pois}(n|b)    \; \prod_{j=1}^{n} f_b(x_j)}
      = e^{-s} \; \left( \frac{s+b}{b} \right)^n \;
        \frac{\prod_{j=1}^{n} f_{s+b}(x_j)}
             {\prod_{j=1}^{n} f_b(x_j)}
\notag \\
q({\bf x}) &\equiv \ln Q({\bf x}) 
           = {-s}  +  \sum_{j=1}^{n} \ln \left( 1+\frac{s f_s(x_j)}
                                                       {b f_b(x_j)}
                                                 \right)
\label{eq:testStatistic}
\end{alignat}
We compute the normalized probability distributions $f(x)$ from the
parton--level matrix elements. This way we construct a log--likelihood
ratio map of all possible final--state phase space configurations
using the normalized probability distributions
$d\sigma(\vec{r})/\sigma_{\rm tot}$ for the signal and background
hypotheses:
\begin{equation}
q(\vec{r}) = -\sigma_{{\rm tot},s} \; {\cal L} \; + \,
             \ln \left( 1 + \frac{d\sigma_s(\vec{r})}
                                 {d\sigma_b(\vec{r})}  
                 \right)
\label{eq:likelimap}
\end{equation}
${\cal L}$ is the integrated luminosity. To construct the
single--event probability distribution $\rho_{1,b}(q)$ we combine the
background event weight with the log--likelihood ratio map
$q(\vec{r})$ from Eq.(\ref{eq:likelimap}), which in general is not
invertable:
\begin{equation}
\rho_{1,b}(q_0) = \int d\vec{r} \; \frac{d\sigma_b(\vec{r})}
                                        {\sigma_{{\rm tot},b}}
                                \; \delta \left( q(\vec{r}) - q_0 \right)
\label{eq:rho_1}
\end{equation} 
For multiple events, the distribution of the log--likelihood ratio
$\rho_{n,b}$ can be computed by repeated convolutions of the single
event distribution. This convolution we can either perform implicitly
with approximate Monte Carlo techniques~\cite{junk}, or analytically
using a Fourier transform~\cite{clfft}.\smallskip

The expected log--likelihood ratio distribution for a background
including Poisson fluctuations in the number of events takes the form
$\rho_b(q) = \sum_n {\rm Pois}(n|b) \times \rho_{n,b}(q)$.  To compute
this $\rho_b(q)$ from the single--event likelihood $\rho_{1,b}(q)$
given by Eq.(\ref{eq:rho_1}) we first Fourier transform all $\rho$
functions into complex--valued functions of the Fourier conjugate of
likelihood ratio, \eg $\overline{\rho_{1,b}}(\overline{q})$. The
Fourier--transformed $n$-event likelihood ratio is now given by
$\overline{\rho_{n,b}} = (\overline{\rho_{1,b}})^n$ equivalent to a
convolution in $q$-space.  The sum over $n$ in the formula for
$\rho_b(q)$ now has a simple form in the Fourier domain:
$\overline{\rho_b} = \exp[b \; (\overline{\rho_{1,b}} - 1)]$.  For the
signal--plus--background hypothesis we expect $s$ events from the
$\rho_{1,s}$ distribution and $b$ events from the $\rho_{1,b}$
distribution. Similar to the above formula we have
$\overline{\rho_{s+b}} = \exp[ b (\overline{\rho_{1,b}} - 1) + s
(\overline{\rho_{1,s}} - 1)]$. This form we can transform back and
obtain the log-likelihood ratio distributions $\rho_b(q)$ and
$\rho_{s+b}(q)$.\medskip

Given a log-likelihood ratio $q$ we can calculate the background-only
confidence level, ${\rm CL}_b$:
\begin{equation}\label{eq:clb}
{\rm CL}_b(q) =\int_{q}^\infty dq' \; \rho_b(q')
\end{equation}
To estimate the discovery potential of a future experiment we assume
the signal--plus--background hypothesis to be true and compute ${\rm
  CL}_b$ for the median of the signal--plus--background distribution
$q^*_{s+b}$.  This expected background confidence level can be
converted into an equivalent number of Gaussian standard deviations
and the significance written as $Z~\sigma$ by implicitly solving ${\rm
  CL}_b(q^*_{s+b}) = \left( 1-{\rm erf}(Z/\sqrt{2} \right)/2$ for $Z$.

\subsection{Higgs Decay to Muons}

\begin{figure}[t]
\includegraphics[width=.45\textwidth]{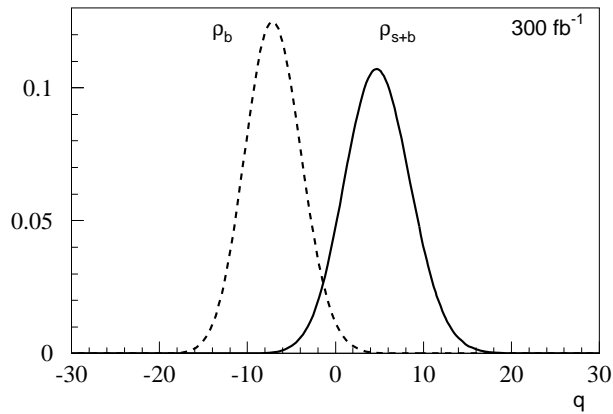}
\caption{Normalized $\rho_{b}(q)$ and $\rho_{s+b}(q)$ distributions, 
  corresponding to the full--experiment log--likelihood ratio in
  Eq.(\ref{eq:testStatistic}). These distributions define the
  expected significance.}
\label{fig:logq}
\end{figure}

To determine the maximal significance in a strict sense we should not
include detector effects which always decrease the significance.
However, in our example of weak--boson--fusion $H\to \mu \mu$ the
experimental resolution on the invariant mass $m_{\mu\mu}$ is much
larger than the Higgs width: about 1.6 $\gev$ for {\sc CMS} and 2.0
$\gev$ for {\sc Atlas}~\cite{massResolution}.  To obtain a
semi--realistic result we introduce a Gaussian smearing for
$m_{\mu\mu}$ into Eq.(\ref{eq:int_first}). This Gaussian shape is just
a simple numerical choice and could be replaced with any other
smearing prescription or fast detector simulation. We convolute our
momentum smearing with the Breit--Wigner--shaped Higgs propagator;
 in our case, the combination is completely dominated
by the much larger Gaussian width.\smallskip

We introduce a new random number $r_m^*$ corresponding to the smeared
$m_{\mu\mu}^*$ and integrate over a transfer function from the true
$m_{\mu\mu}$ to the smeared $m_{\mu\mu}^*$ by aligning one of the
original random numbers $r_m$ with $m_{\mu\mu}$:
\begin{equation}
\sigma_{\rm tot} 
= \int d \vec{r}_\perp d r_m^* 
  \int_{-\infty}^{\infty} d r_m 
  \; M({\vec{r}}) \; d\sigma({\vec{r}}) \; W(r_m,r_m^*)
\label{eq:int_m}
\end{equation}
The original random number vector $\vec{r}$ is split into $\vec{r} =
\{\vec{r}_\perp,r_m\}$. In our case, the transfer function $W$
is a normalized Gaussian giving the likelihood to reconstruct
$m_{\mu\mu}^*$ given the true $m_{\mu\mu}$ and the experimental mass
resolution. We trivially get back Eq.(\ref{eq:int_first}) for
$W(r_m,r_m^*) \to \delta(r_m-r_m^*)$.

In general one must be careful about the mapping between parton--level
quantities and their observable counter-parts.  As in most
experimental analyses that use the matrix element method, the jet
direction is assumed to be well-measured.  We do not include a jet
energy scale in the transfer function $W$, because, unlike the
top-mass measurement with a hadronically decaying resonance, the jet
energy scale is not a dominant experimental issue for this search.  In
particular, the jet momenta have relatively flat distributions, \ie
their variation on the scale of detector effects is
small~\cite{wbf_muon}.  In the general case, one should consider all
permutations between out-going quarks and gluons with jets; however,
in the case of weak boson fusion, the signal-like regions of phase
space have more than three units of pseudorapidity separating the
tagging jets, which makes the association of parton to jet
unambiguous. In other words, adding the alternative jet--parton
assignment would give a negligible contribution to the event weight.
The correspondence of the muons is also clear due to their charge.
\medskip

From Eq.(\ref{eq:int_first}) it is obvious how to include an
experimental mass resolution: we replace the event weights $(M \,
d\sigma)$ by the integral $ \left( M\int d r_m \, d\sigma \, W
\right)$ and evaluate them over the smeared phase space $\{
\vec{r}_\perp, r_m^*\}$. Because the random numbers form a (minimal)
basis for all final state configurations there is no `back door' for the
true (infinitely well measured) $m_{\mu \mu}$ to enter the
likelihood calculation. A rough approximation to incorporating the
$m_{\mu\mu}$ mass resolution could be an increased physical Higgs
width. It replaces the Gaussian smearing with a Breit--Wigner
function; we compared this approximate method with the proper smearing procedure
and found that the difference in the final results was small but not
negligible.\medskip

For all details of the signal and background simulation (using {\sc
  CTEQ 5L} parton distributions) we refer to Ref.~\cite{wbf_muon}.
There, after very basic cuts the signal cross section for a $120\gev$
Higgs is $0.22\fb$, hidden under $0.33\fb$ of electroweak $Z$
production and $2.6\fb$ of QCD $Z$ production, where the $Z$ decays
into muons. All other backgrounds combined contribute less than
$0.01\fb$, which allows us to neglect them.  It is worth mentioning
that the electroweak $Z$ production consists of as many as 48 diagrams
for a fixed flavor configuration, which is substantially more
complicated than the search for single top
production~\cite{singleTop}. Conservatively assuming no additional
information from higher--order jet radiation, we could apply $K$
factors to the signal and cross section rates, but this would lead
beyond this proof--of--principle letter. \smallskip

To probe the likelihood ratio over the full phase space, we relax the
cuts for a $120\gev$ Standard Model Higgs to mere acceptance cuts. All
cross sections are finite, so the cut values have no effect on the
likelihood we obtain. Using $2^{20}$ points we integrate over the
final--state phase space projected onto the log--likelihood ratio
$q(\vec{r})$ according to Eq.(\ref{eq:rho_1}).  The phase space points
used for this integration are defined by the same grid we use for the
integration over the signal and background amplitudes described in
Eq.(\ref{eq:int_m}); this way we can check the total rates to ensure
that the likelihood integration covers the entire phase space. For
each phase space point we integrate over the true $m_{\mu\mu}$ as
shown in Eq.(\ref{eq:int_m}), using a proper phase space mapping. Note
that this internal integration does not have to use the same grid for
signal and background.\medskip

The resulting log-likelihood distributions $\rho_b(q)$ and
$\rho_{s+b}(q)$ are shown in Fig.~\ref{fig:logq}. From the background
pdf we extract the signal significance for an integrated luminosity of
$300\fb^{-1}$ as $3.54~\sigma$ for {\sc CMS} and $3.19~\sigma$ for
{\sc Atlas}.  Note that this significance estimate neglects
theoretical uncertainty in the overall rate since the signal has small
higher-order corrections~\cite{wbf_qcd} and the normalization of the
background will be well measured with $300\fb^{-1}$ of data.  Also
note that this significance does not include a minijet veto because
only two jets are included in our parton--level transition amplitude;
in principle, the same procedure could be repeated with a higher-order
tree-level or a next-to-leading order calculation. Following
Ref.~\cite{wbf_muon} we can estimate the effect of a minijet veto,
which increases the significance to $\sim4.4~\sigma$ for {\sc CMS}.
Survival probabilities for the veto neglect pile-up effects, which
will degrade the enhancement in significance.  Combining both
experiments the significance even without a minijet veto is
$4.77~\sigma$.

\begin{figure}[t]
\includegraphics[width=.45\textwidth]{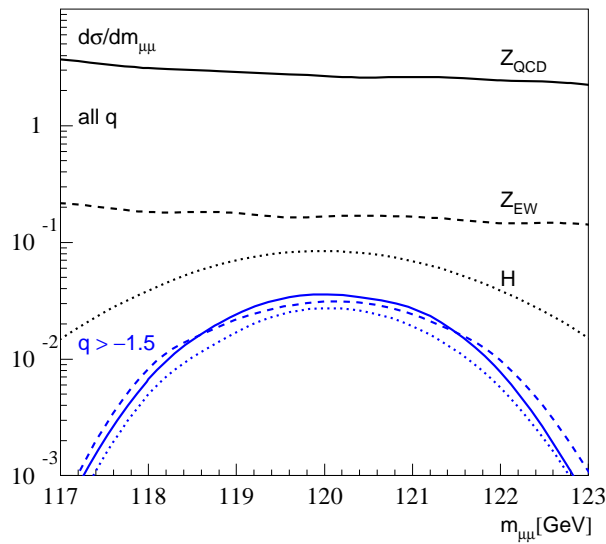}
\caption{Muon invariant mass distribution for the 
  120\gev\ Higgs signal and $Z$+jets background with acceptance cuts
  only (upper curves) and after a cut on the log--likelihood ratio
  $q(\vec{r})>-1.5$ (lower curves).  The curves correspond to {\sc
    CMS} and illustrate that events with high $q(\vec{r})$ have an
  increased signal purity and signal--like characteristics.}
\label{fig:mumu}
\end{figure}

The most relevant kinematic distribution is the reconstructed Higgs
mass $m_{\mu\mu}$. In the upper curves of Fig.~\ref{fig:mumu} we show
it for signal and backgrounds without kinematic or likelihood cuts.
The signal shows a smeared mass peak, while the backgrounds are flat.
To illustrate how the method isolates signal--rich phase space
regions, we apply a likelihood ratio cut $q(\vec{r})>-1.5$. Roughly a
third of the signal events survive this cut, and each of the
backgrounds are reduced to a rate comparable to the signal.  After the
likelihood cut the backgrounds show the same kinematic features as the
signal, \ie a peak in $m_{\mu\mu}$.

\subsection{Detector Effects and Reducible Backgrounds}

The procedure for incorporating detector smearing on observables
described above is tailored for smearing of a few observables, which
are isolated in the phase space integration.  Nevertheless, it is
possible to generalize the smearing procedure. In essence, a complete
detector smearing requires an integration over a fixed set of
experimental observables with a nested integration over the remaining
degrees of freedom in the phase space. The latter include the
unsmeared (true) observables, as shown in Eq.(\ref{eq:int_m}), as well
as the unobservable longitudinal component of neutrino momenta at a
hadron collider or the momentum of particles not passing the
acceptance cuts.  As mentioned in the previous section and discussed
in the literature related to the matrix element method, one should
take care to include in the transfer function all relevant detector
effects and consider all permutations that arise from ambiguities in
the mapping from parton--level quantities to their final state
observables.

We usually include detector effects by smearing all final state
four-momenta; however, this can be computationally inefficient. If we
instead choose not to smear some of the observables, we must remain
vigilant to insure that there is no `back door' through which
four-momentum conservation together with unsmeared observables
implicitly evade smearing.  We avoid this `back door' explicitly in
Eq.(\ref{eq:int_m}) by factorizing the basis of the phase space into
orthogonal components $r_m$ and $r_\perp$.  \smallskip

After generalizing our method to smear multiple observables we can now
incorporate reducible backgrounds, \ie background whose final--state
configurations have more degrees of freedom than the signal. We simply
pick a set of observables that is common to all signal and background
processes, and marginalize the additional background degrees of
freedom. Flavor tagging efficiencies and fake rates can be included in
the event weights through $W$. In these scenarios, the interpretation
of the resulting significance is more vague: it is the maximal
significance given the specified set of observables and the
assumptions in the transfer and measurement functions.

\subsection{Conclusions} 

We have described a way to compute the mathematically strict maximum
significance for a set of signal and background processes at the
parton--level. Our method is based on the Neyman--Pearson lemma and
can be used to decide if a new physics search at high--energy
colliders has a sufficiently large discovery potential to justify a
dedicated analysis.  \smallskip

While our example is fairly simple, including only irreducible
backgrounds and incorporating experimental resolution for only a
single observable, we have outlined the extension of the method to
include general detector effects.  This approach to including detector
effects follows closely the recent experimental work at the Tevatron
referred to as `the matrix element method'.  The next step will be to
implement this likelihood computation into a parton--level event
generator with a simple and fast simulation of detector
effects~\cite{whizard}.\smallskip

Weak--boson--fusion production of a Higgs boson with a subsequent
decay to muons is the perfect showcase for this new method: it suffers
from very low signal rate and from the lack of distributions that
clearly distinguish signal from background.  A very basic cut analysis
in Ref.~\cite{wbf_muon} quotes a significance of $1.8~\sigma$ for
$300\fb^{-1}$ for a single experiment. In particular,
Ref.~\cite{wbf_muon} found that a cut analysis was likely not the
best--suited strategy for this signal. Applying our method we arrive
at a possible maximum significance of $3.54~\sigma$ ({\sc CMS} with
$300\fb^{-1}$).  By increasing the complexity of the final state,
higher--order QCD effects can be exploited using a minijet
veto~\cite{wbf_minijet}, which could increase the significance to
$\sim 4.4~\sigma$. Not only is this result grounds for a more careful
study by the experimental collaborations, but it also indicates that
without a luminosity upgrade {\sc Atlas} and {\sc CMS} combined may be
able to observe the decay $H \to \mu\mu$.  \bigskip \bigskip

\begin{acknowledgments}
  We would like to thank the Aspen Center of Physics and the theory
  group of the MPI for Physics for their generous hospitality.
  Moreover, we would like to thank Dave Rainwater for his help with
  our example process, Daniel Whiteson for his useful comments on our
  draft and Markus Diehl for his explanations on optimal observables.
\end{acknowledgments}
 

\end{document}